\documentclass[epj]{svjour}

\usepackage{graphicx}
\usepackage{hyperref}
\usepackage[numbers,sort&compress]{natbib} 
\usepackage{amsmath}
\usepackage{amssymb}
\usepackage{placeins}

\newcommand{\de}{\delta}
\newcommand{\eref}[1]{Eq.~(\ref{#1})}
\newcommand{\tref}[1]{Tab.~\ref{#1}}
\newcommand{\fref}[1]{Fig.~\ref{#1}}

\pdfoutput=1

\allowdisplaybreaks

\begin{document}

\title{On non-primitively divergent vertices of Yang-Mills theory}
\author{Markus. Q. Huber
\thanks{\email{markus.huber@uni-graz.at}}
}

\institute{Institute of Physics, University of Graz, NAWI Graz, Universit\"atsplatz 5, 8010 Graz, Austria}

\date{\today}

\abstract{
Two correlation functions of Yang-Mills beyond the primitively divergent ones, the two-ghost-two-gluon and the four-ghost vertices, are calculated and their influence on lower vertices is examined.
Their full (transverse) tensor structure is taken into account.
As input, a solution of the full two-point equations - including two-loop terms - is used that respects the resummed perturbative ultraviolet behavior.
A clear hierarchy is found with regard to the color structure that reduces the number of relevant dressing functions.
The impact of the two-ghost-two-gluon vertex on the three-gluon vertex is negligible, which is explained by the fact that all non-small dressing functions drop out due to their color factors.
Only in the ghost-gluon vertex a small net effect below $2\%$ is seen.
The four-ghost vertex is found to be extremely small in general.
Since these two four-point functions do not enter into the propagator equations, these findings establish their small overall effect on lower correlation functions.
} 

\maketitle

\section{Introduction}
\label{sec:introduction}

Nonperturbative methods are indispensable to account for the strongly coupled nature of quantum chromodynamics (QCD) in the low momentum regime.
Depending on the method, different challenges have to be overcome, while certain aspects are particularly simple in one or the other approach.
In the case of functional methods, an advantage is clearly that they are a continuum approach and large scale separations can be handled.
In addition, the quark mass is a parameter that can be tuned at will and the chiral limit is easily accessible what sets it apart from lattice calculations.
However, realizing any quantitative calculation requires a truncation of an infinitely large system of equations.

These truncations are the systematic error of this meth\-od which is unfortunately difficult to estimate.
While results can be improved using the freedom implied by the use of models for some quantities, it would of course be ideal to realize a self-contained calculation where the strong coupling constant and the quark masses are the only external parameters.
Low order truncations do not allow for this possibility and at least three-point functions need to be included.
Recent progress in this direction was achieved with equations of motion in three-dimensional Yang-Mills theory \cite{Huber:2016tvc} and the functional renormalization group (FRG) for Yang-Mills theory \cite{Cyrol:2016tym} and QCD \cite{Cyrol:2017ewj}.
The groundwork was laid by various calculations of vertices, e.g., \cite{Schleifenbaum:2004id,Kellermann:2008iw,Alkofer:2008dt,Huber:2012zj,Huber:2012kd,Hopfer:2013np,Aguilar:2013xqa,Blum:2014gna,Binosi:2014kka,Williams:2014iea,Cyrol:2014kca,Eichmann:2014xya,Mitter:2014wpa,Williams:2015cvx}.
In the case of Yang-Mills theory, all primitively divergent correlation functions \cite{Huber:2016tvc,Cyrol:2016tym} have been calculated as a self-contained system, viz., the gluon and ghost propagators, the ghost-gluon vertex, the three-gluon vertex and the four-gluon vertex.
Although the agreement with corresponding lattice results \cite{Cucchieri:2007md,Cucchieri:2008fc,Sternbeck:2007ug,Bogolubsky:2009dc,Cucchieri:2011ig,Bornyakov:2013ysa,Maas:2014xma,Cucchieri:2016jwg} was quite good, the question of higher correlation functions needs to be addressed to learn about the convergence properties of functional systems of equations.

All the previously mentioned studies were done in the Landau gauge.
Also in this work this gauge is used as it is the most convenient and consequently best studied one.
Beyond the Landau gauge, only in the Coulomb gauge functional calculations at the level of vertices have been performed \cite{Huber:2014isa,Campagnari:2011bk,Campagnari:2010wc}.
In the generalization of the Landau gauge to linear covariant gauges, on the other hand, only two-point functions have been studied up to now \cite{Aguilar:2015nqa,Huber:2015ria}.
In that case, the inclusion of the longitudinal part of the correlation functions does not only increase the number dressings to be calculated but also increases the numerical complexity.

In this work the two lowest correlation functions beyond the set mentioned above, the two-ghost-two-gluon vertex and the four-ghost vertex, are studied within the Dyson-Schwinger formalism.
The emphasis will be put on the former, because it appears in the equation for the three-gluon vertex and allows completing the equation of the ghost-gluon vertex so that no truncation is required.
Since there is no guidance for their tensor structure from a tree-level tensor, a full basis is constructed.

For this study, the propagator equations are not treated dynamically.
They are solved by themselves and the results are used as fixed input.
Thus, no back-coupling effects of the modifications of the vertices are contained and the studied effects can be purely attributed to truncation effects at the level of three- and four-point functions.
However, the employed setup for the propagators is new insofar as it uses the full, untruncated Dyson-Schwinger equations (DSEs).
The inclusion of the two-loop terms in the gluon propagator DSE allows for the first time to obtain the resummed perturbative behavior at one loop without modifications in the employed vertex models.

The article is organized as follows: Sec.~\ref{sec:props} contains some details on how the propagator results were obtained that are used as input in the rest of this work.
Details on the two four-point functions are presented in Sec.~\ref{sec:4p_funcs} and the corresponding results are presented in Sec.~\ref{sec:results}.
Conclusions are drawn in Sec.~\ref{sec:conclusions}.
Some details on calculations of color tensors are deferred to an appendix.

\section{The propagator equations}
\label{sec:props}

The three- and four-point functions will be solved self-consistently using fixed input from the propagators.
Since the employed setup for the propagator equations contains some new features, it will shortly be explained here; details will be given elsewhere \cite{Huber:2017ip}.
Besides two-loop diagrams it also includes a new approach to the treatment of the resummed perturbative behavior in the UV.

\begin{figure}[tb]
  \includegraphics[width=0.35\textwidth]{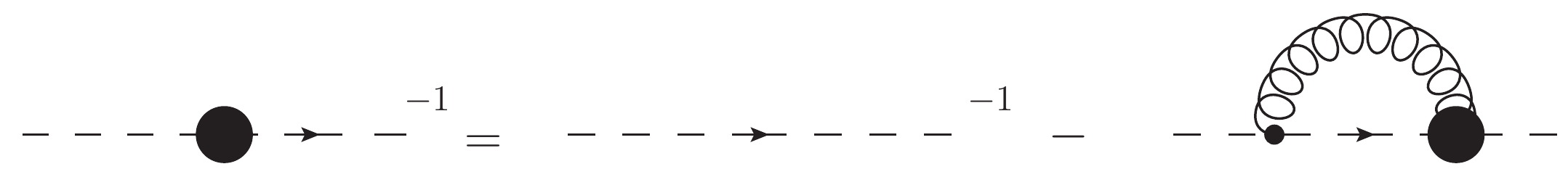}
  \vskip5mm
  \includegraphics[width=0.45\textwidth]{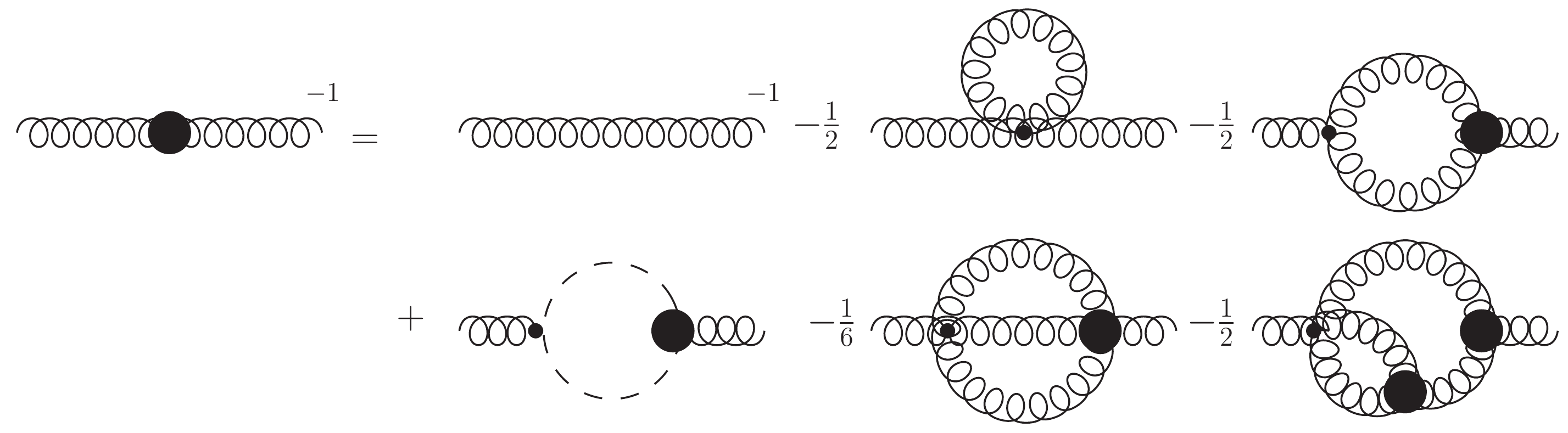}
   \begin{center}
  \caption{\label{fig:gh-gl-DSEs}Top: The ghost propagator DSE. Here and in other figures, internal propagators are dressed, and thick blobs denote dressed vertices, wiggly lines gluons, and dashed ones ghosts. Bottom: The gluon propagator DSE. The loop diagrams are called tadpole, gluon loop, ghost loop, sunset, and squint.
  Feynman diagrams were created with Jaxodraw \cite{Binosi:2003yf}.}
 \end{center}
\end{figure}

The propagator DSEs depend only the primitively divergent correlation functions, see \fref{fig:gh-gl-DSEs}: The ghost-gluon, the three-gluon and the four-gluon vertices.
The latter appears only in the sunset diagram, which was found to be small in all previous calculations \cite{Mader:2013ru,Meyers:2014iwa,Hopfer:2014th,Huber:2016tvc}.
However, only the tree-level tensor structure has been taken into account so far.
The ghost-gluon and the three-gluon vertices, on the other hand, are known to be quantitatively important, see, e.g., \cite{Huber:2012kd,Aguilar:2013xqa,Aguilar:2013vaa,Blum:2014gna,Huber:2016tvc}.

\begin{figure*}[tb]
 \includegraphics[height=2.5cm]{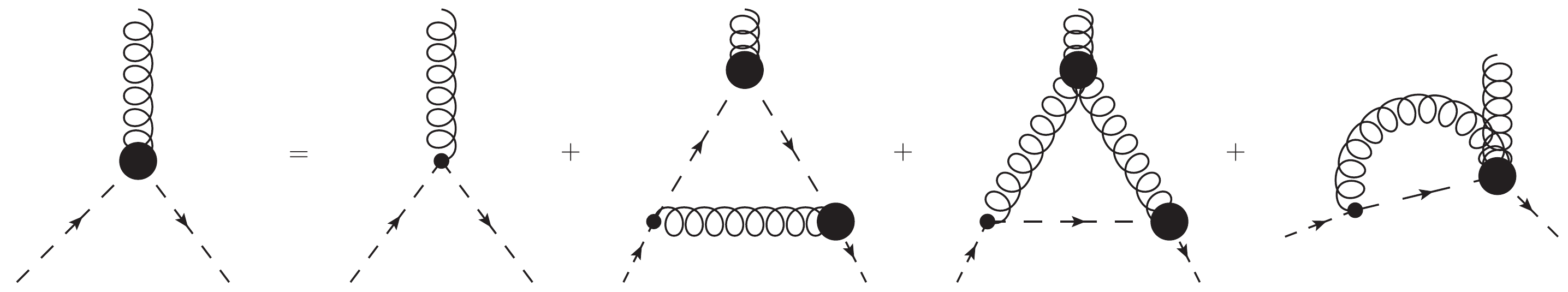}\\
 \vskip5mm
 \includegraphics[height=2.5cm]{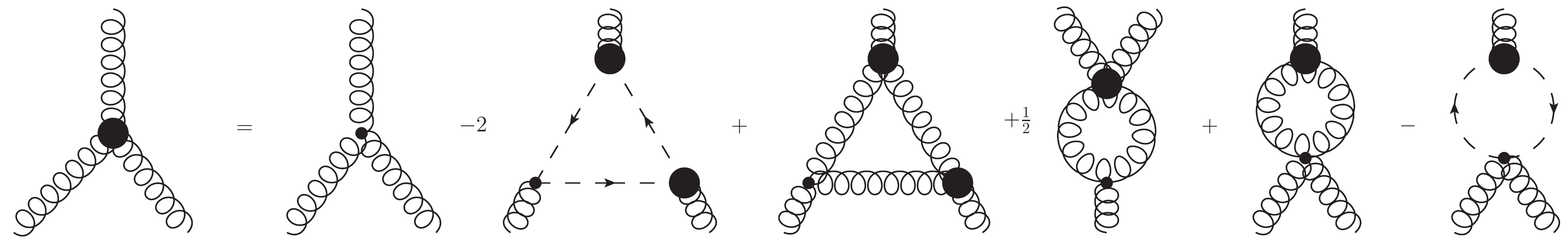}\\
 \vskip5mm
 \includegraphics[height=5cm]{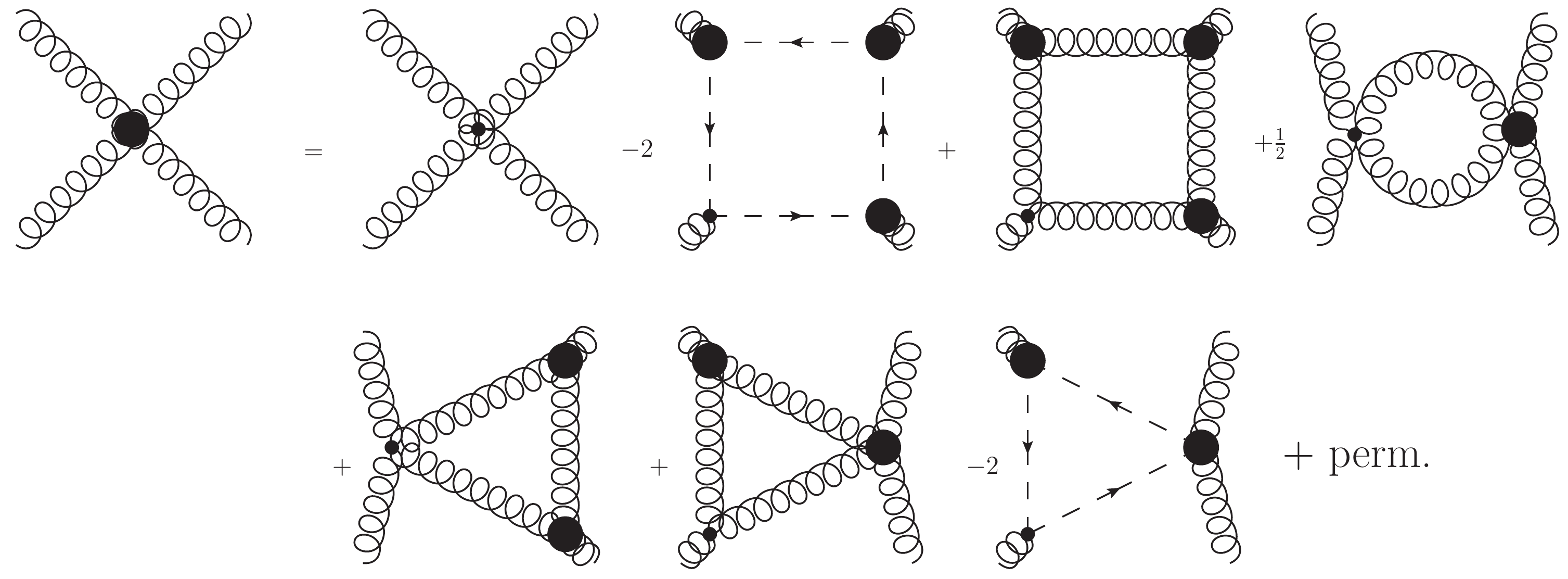}
 \begin{center}
 \caption{Top: The untruncated $c$-DSE of the ghost-gluon vertex where the first derivative was done with respect to the ghost field.
 Center: One-loop truncated DSE of the three-gluon vertex.
 Since the equation is symmetrized in the calculation, two swordfish diagrams can be merged, leading to a factor of $1$ in front instead of the symmetry factor $1/2$.
 Bottom: One-loop truncated DSE of the four-gluon vertex.}
 \end{center}
 \label{fig:ghg-3g-4g_DSEs}
\end{figure*}

\begin{figure*}[tb]
  \includegraphics[width=0.45\textwidth]{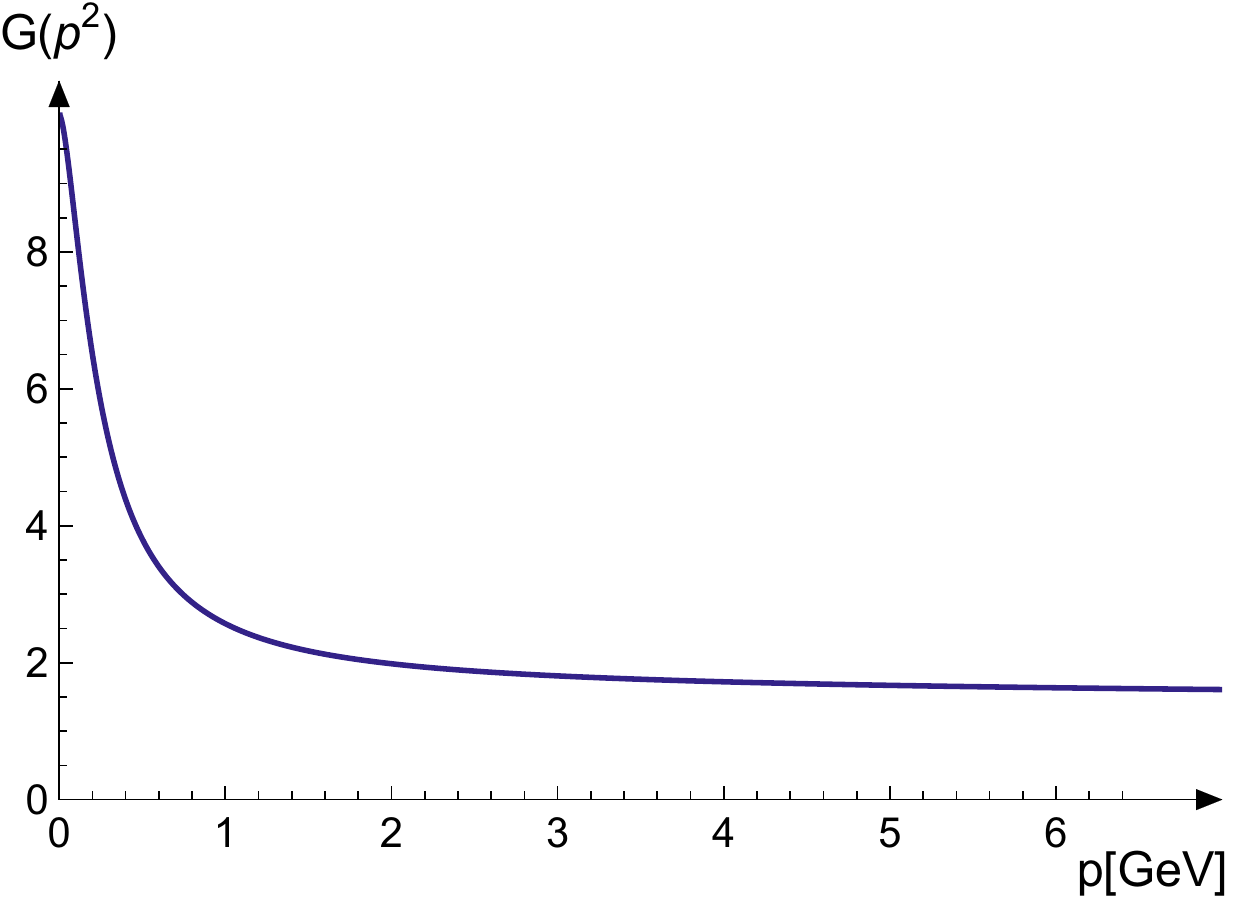}\hfill
  \includegraphics[width=0.45\textwidth]{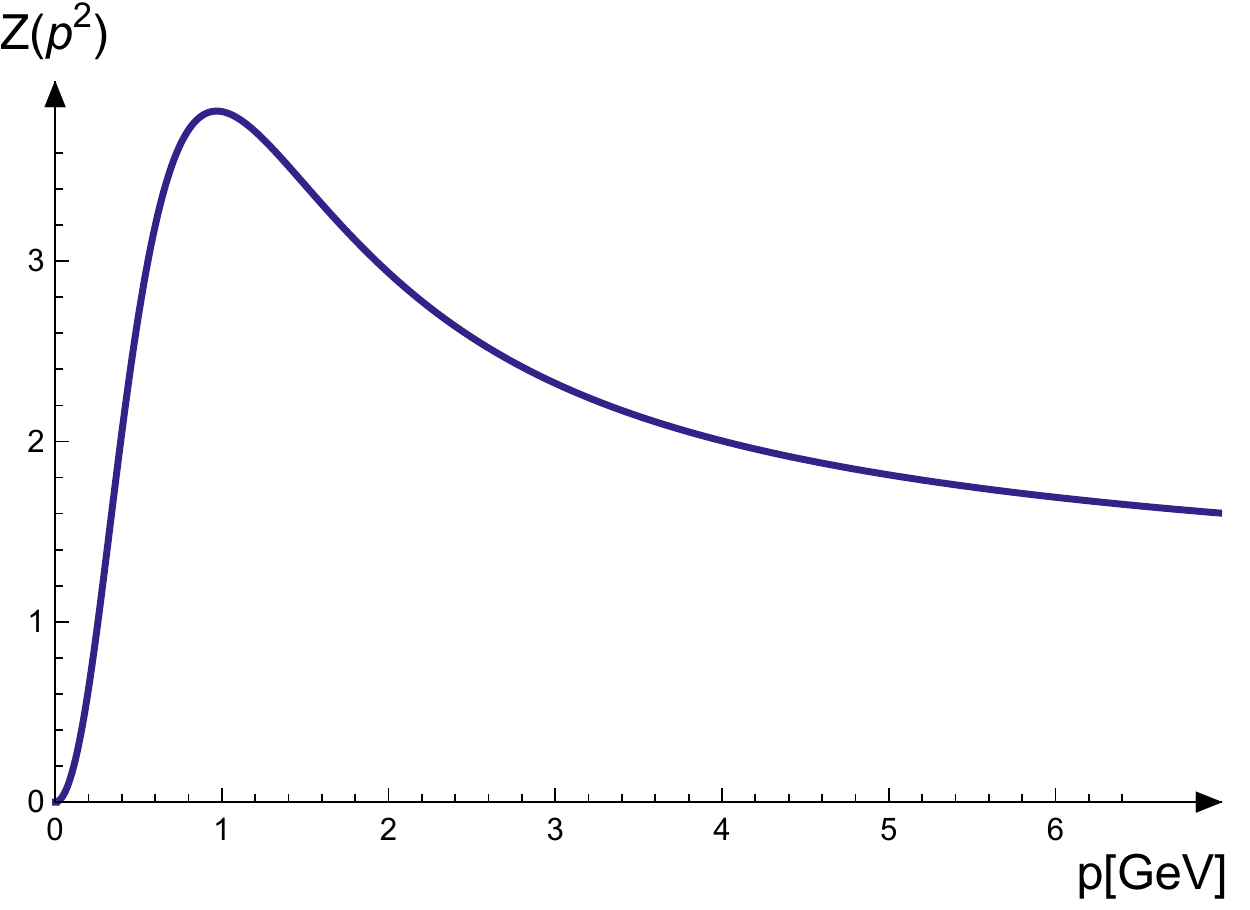}\\
  \includegraphics[width=0.45\textwidth]{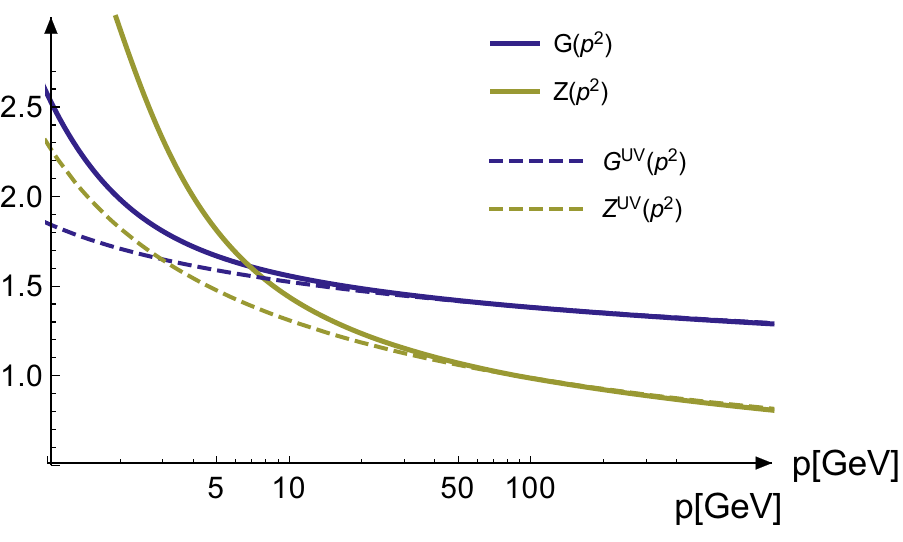}\hfill
  \includegraphics[width=0.45\textwidth]{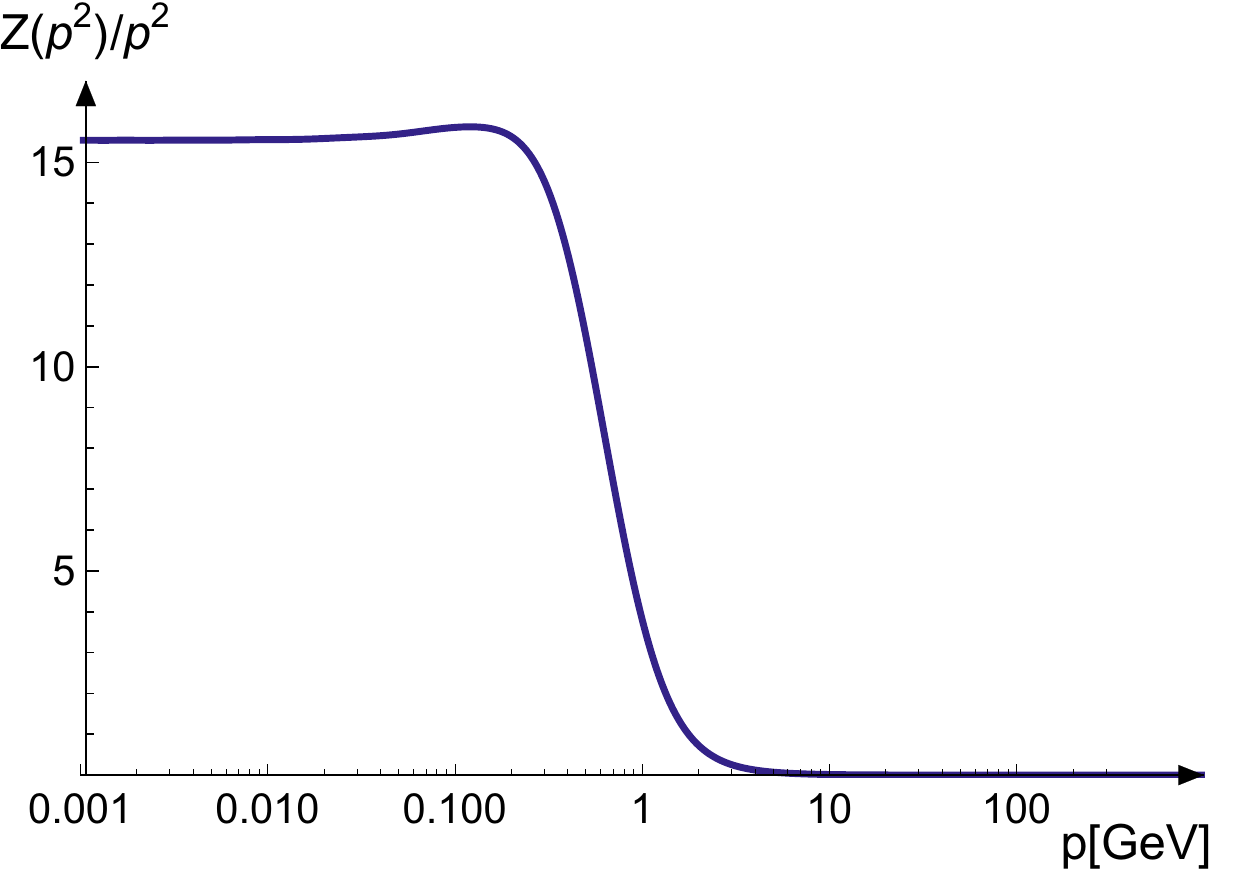}
  \begin{center}
  \caption{\label{fig:res_props}Top: Ghost (left) and gluon (right) dressing functions.
  Bottom: Gluon propagator (right) and the UV behavior of the propagator dressing functions compared to the one-loop resummed behavior (left).}
 \end{center}
\end{figure*}

The gluon propagator DSE is solved including the two-loop diagrams.
This allows not only including important quantitative contributions, but also obtaining the resum\-med one-loop behavior in the UV.
Without the two-loop terms, crucial contributions to the resummed series are missing.
To account for this, an additional term can be added to the gluon loop diagram \cite{vonSmekal:1997vx,Fischer:2002eq,Huber:2012kd}.
At the same time, the renormalization constant of the three-gluon vertex, $Z_1$, in this diagram is dropped.
This can also be interpreted as making $Z_1$ momentum dependent.
However, when the two-loop diagrams are included, this procedure is not only no longer necessary but would even spoil the calculation.
Thus, $Z_1$ and $Z_4$ are included with their values from the Slavnov-Taylor identities (STIs):
\begin{align}\label{eq:Z1Z4}
 Z_1=\frac{Z_3}{\widetilde Z_3}, \qquad Z_4=\frac{Z_3}{\widetilde Z_3^2}.
\end{align}
Here, the renormalization constant of the ghost-gluon vertex, $\widetilde Z_1$, was set to $1$, as this vertex is finite in the Landau gauge.
Thus, this problem does not arise for diagrams with a bare ghost-gluon vertex and only the gluon loop needs to be treated specially in a one-loop truncation of the propagator equations.

It remains to specify how the vertices are included.
Since primarily we aim at qualitative understanding of the importance of higher correlation functions, a minimal setup is chosen for the input: The four-gluon vertex is taken as bare and for the three-gluon vertex the following model dressing for the tree-level tensor is used:
\begin{align}
 D^{AAA}&(p^2,q^2,r^2)=\frac{G(\overline{p}^2)}{Z(\overline{p}^2)}\frac{\overline{p}^2}{\overline{p}^2+\Lambda_s^2}.
\end{align}
$\overline{p}^2$ is given by $(p^2+q^2+r^2)/2$ and $G(p^2)$ and $Z(p^2)$ are the dressing functions of the ghost and gluon propagators, respectively.
A perturbative one-loop analysis \cite{Fischer:2002eq,Huber:2012kd} of the gluon propagator DSE, shows that the vertex should be proportional to $G(p^2)/Z(p^2)$ as expected also from its STI.
However, since this term is IR divergent, a damping function is added to tame the UV part in the IR.
The damping scale is chosen as $\Lambda^2_s= 1.54\,\text{GeV}^2$.
Since we are interested mainly in the correct UV behavior, no IR contribution is added as in Ref.~\cite{Huber:2012kd}.

The ghost-gluon vertex is coupled dynamically using the one-momentum approximation explained in Sec.~\ref{sec:results}.
The employed DSE is the one shown in \fref{fig:ghg-3g-4g_DSEs} but with the two-ghost-two-gluon vertex set to zero.
It turned out to be necessary to use a dressed ghost-gluon vertex, since with a bare vertex the bump in the gluon dressing function was found to be so large that problems with convergence of the three-gluon vertex equation appeared \cite{Blum:2014gna,Eichmann:2014xya}.
Interestingly, comparing the results obtained with a bare and a dressed ghost-gluon vertex, the difference for the propagators is quite small but, as it turned out, decisive.
In addition, using numeric results for the ghost-gluon vertex ensures that all perturbative terms up to two-loop are contained in the propagator equations.
For the three-gluon vertex this is ensured via the correct running of the model.
For the four-gluon vertex the bare one is sufficient, since it only appears in a two-loop diagram.

To solve the propagator equations, a momentum subtraction scheme is employed.
The ghost propagator DSE is subtracted at vanishing momentum, while the gluon propagator DSE is renormalized in the perturbative regime.
The specific renormalization conditions are $G(0)=10$ and $Z(x_s)=1$, with $x_s=7720\,\text{GeV}^2$.
To deal with quadratic divergences arising in the gluon propagator DSE \cite{Huber:2014tva}, a second renormalization condition $D(0)=15.54\,\text{GeV}^{-2}$ is introduced that fixes the gluon propagator $D(p^2)$ at zero momentum.
As renormalized coupling $\alpha(s)=0.05$ is chosen.
This fixes indirectly the value for the scale $s$.
Although in principle a range of values can be chosen for $\alpha(s)$, it is  advantageous to choose a small value to recover resummed perturbation theory with only one- and two-loop terms.

In the renormalization scheme, known as \emph{MiniMOM} scheme \cite{vonSmekal:1997vx,vonSmekal:2009ae}, the renormalization constants of the ghost and the gluon propagators, $\widetilde Z_3$ and $Z_3$, respectively, are fixed implicitly.
Only once a solution is obtained, they can be calculated.
A useful test is to calculate them for various momentum points and check their momentum independence.
However, the renormalization constants of the three- and four-gluon vertices, $Z_1$ and $Z_4$, respectively, are required.
To obtain a self-consistent solution, the equations are solved with fixed $Z_1$ and $Z_4$ as calculated from the corresponding STIs, \eref{eq:Z1Z4}.
From this solution, $\widetilde{Z}_3$ and $Z_3$ and in turn $Z_1$ and $Z_4$ are calculated.
Then the equations are solved again until $Z_1$ and $Z_4$ have converged.
This method requires to start the iteration process from a solution obtained somehow else to provide access to $Z_1$ and $Z_4$.
For this purpose, the one-loop equation was solved with the usual replacement of $Z_1$.
The ghost and gluon dressing functions obtained with this setup are shown in \fref{fig:res_props}, where also the UV behavior is explicitly compared to one-loop resummed perturbation theory.
To set the scale, the maximum of the gluon dressing function is moved to $p^2=0.94\,\text{GeV}^2$ which is the value extracted from lattice calculations \cite{Sternbeck:2006rd,Bogolubsky:2009dc}.

\section{The two-ghost-two-gluon and four-ghost vertices}
\label{sec:4p_funcs}

Although the two-ghost-two-gluon vertex depends on three independent momenta, its complexity in Lorentz space is not that drastic.
The reason is that only the transverse part is needed which consists of only five tensors.
It is convenient to combine them into tensors symmetric or antisymmetric under exchange of the gluon momenta $p$ and $q$:
\begin{subequations}
\begin{align}
 \tau^1_{\mu\nu}(p,q;r,s)&= t_{\mu\nu}(p,q),\\
 \tau^2_{\mu\nu}(p,q;r,s)&= t_{\mu\alpha}(p,p)t_{\alpha\nu}(r,q) + t_{\mu\alpha}(p,r)t_{\alpha\nu}(q,q),\\
 \tau^3_{\mu\nu}(p,q;r,s)&= t_{\mu\alpha}(p,p)t_{\alpha\nu}(r,q) - t_{\mu\alpha}(p,r)t_{\alpha\nu}(q,q),\\
 \tau^4_{\mu\nu}(p,q;r,s)&= t_{\mu\alpha}(p,p)t_{\alpha\nu}(q,q),\\
 \tau^5_{\mu\nu}(p,q;r,s)&= t_{\mu\alpha}(p,r)t_{\alpha\nu}(r,q),
\end{align}
\end{subequations}
with $t_{\mu\nu}(p,q)=g_{\mu\nu} p\cdot q -p_\mu q_\nu$. $r$ and $s$ are the antighost and ghost momenta, respectively.

\begin{table}
 \begin{center}
 \begin{tabular}{c|c|c|c|c|c|c|c|c}
  \hline
   & $\sigma_1$ & $\sigma_2$ & $\sigma_3$ & $\sigma_4$ & $\sigma_5$ & $\sigma_6$ & $\sigma_7$ & $\sigma_8$\\
   \hline\hline
  $a\leftrightarrow b$ & + & + & + & - & - & - & - & +\\
  \hline
  $c\leftrightarrow d$ & + & + & + & - & - & + & - & -\\
  \hline
 \end{tabular}
 \caption{Symmetry properties of the color basis tensors for four-point functions given in \eref{eq:color_basis_4p}.}
 \label{tab:color_tensors_symmetries}
 \end{center}
\end{table}

\begin{figure*}[tb]
 \begin{center}
  \includegraphics[width=0.9\textwidth]{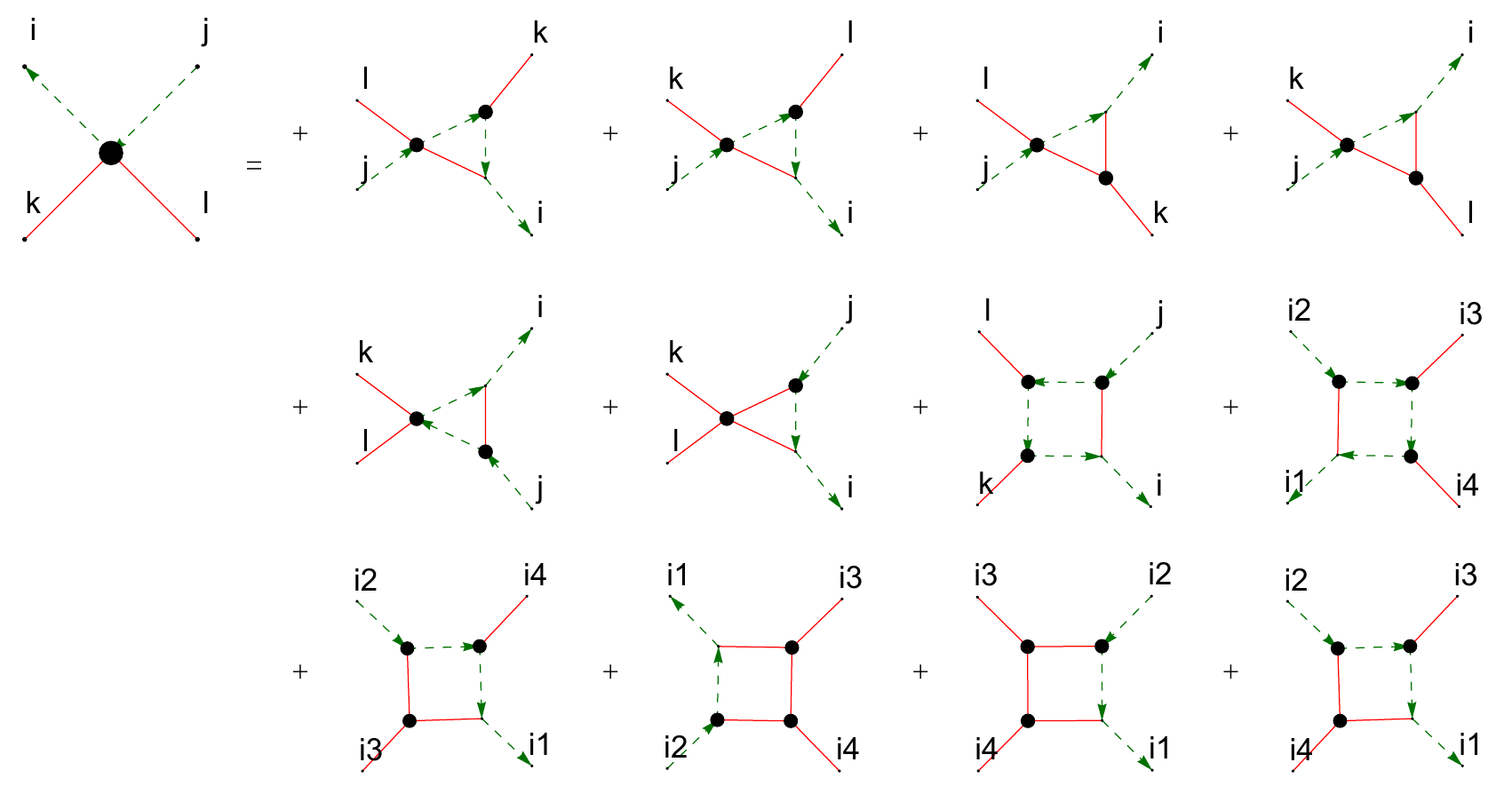}
 \end{center}
 \caption{The truncated $c$-DSE of the two-ghost-two-gluon vertex where the first derivative was done with respect to the antighost field.
 Continuous red lines denote gluons, dashed green lines ghosts.
 The truncation discards only one diagram involving a two-ghost-three-gluon vertex.
 }
 \label{fig:2gh2gl_DSE}
\end{figure*}

\begin{figure*}[tb]
 \begin{center}
  \includegraphics[width=0.9\textwidth]{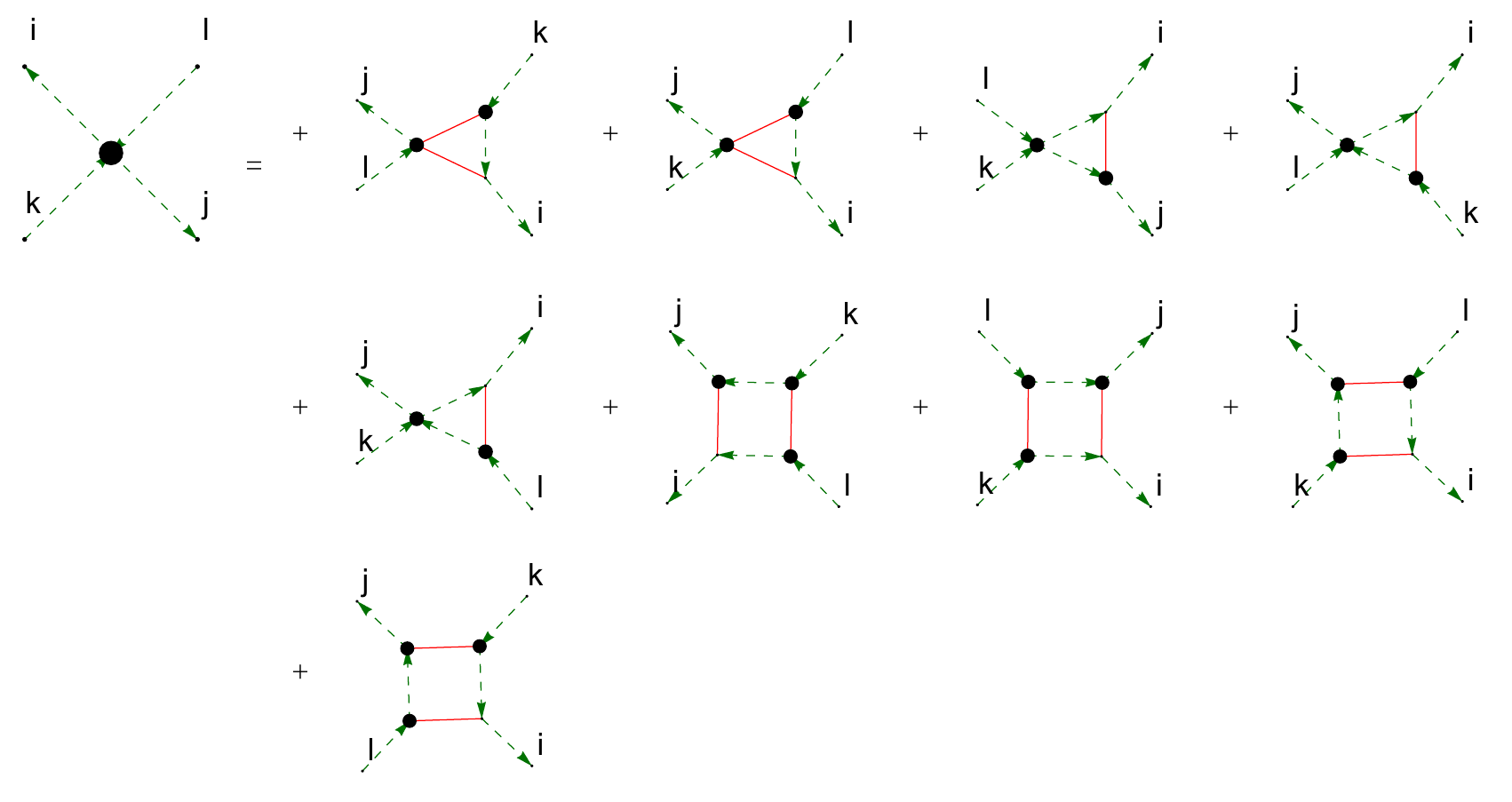}
 \end{center}
 \caption{The truncated DSE of the four-ghost vertex.
 The truncation discards only one diagram involving a four-ghost-one-gluon vertex.}
 \label{fig:4gh_DSE}
\end{figure*}

In color space, there are 15 possible tensors:
\begin{align*}
 \de\, \de&:& 3\,\text{combinations}\\
 f\, f&:& 3\,\text{combinations}\\
 d\,d&:& 3\,\text{combinations}\\
 d\,f&:& 6\,\text{combinations}
\end{align*}
$f$ and $d$ are the antisymmetric and symmetric structure constants of $SU(N)$, and $\de$ is the Kronecker delta.
For $SU(2)$, the symmetric structure constant vanishes and only 6 tensors can be constructed.
There are 6 identities that reduce the number from 15 to 9 independent tensors \cite{Pascual:1980yu}:
\begin{align}\label{eq:SUN_id_fd}
 &f^{abo}d^{cdo}+f^{aco}d^{bdo}+f^{ado}d^{aco}=0 \text{ and 2 permutations},\\
 \label{eq:SUN_id_ff}
 &f^{abo}f^{cdo} = \frac{2}{N}\left(\de^{ac}\de^{bd}-\de^{ad}\de^{cb}\right)+d^{aco}d^{bdo}-d^{ado}d^{cbo}\\
  &\qquad\qquad \text{ and 2 permutations.}\nonumber
\end{align}
The Jacobi identity is a combination of the permutations of \eref{eq:SUN_id_ff}.
Finally, for $SU(3)$ an additional identity reduces the number of independent tensors to 8:
\begin{align}
 \de^{ab}&\de^{cd}+\de^{ac}\de^{bd}+\de^{ad}\de^{bc} =\\
 &3(d^{abo}d^{cdo}+d^{aco}d^{bdo}+d^{ado}d^{bco}).
\end{align}
A full basis can be chosen thus as
\begin{subequations}
\begin{align}
 \overline\sigma_1^{abcd}& = f^{acd}f^{bde}, \\
 \overline\sigma_2^{abcd}& = \de^{ab}\de^{cd},\\
 \overline\sigma_3^{abcd}& = \de^{ac}\de^{bd}, \\
 \overline\sigma_4^{abcd}& = \de^{ad}\de^{bc}, \\
 \overline\sigma_5^{abcd}& = f^{abe}f^{cde},\\
 \overline\sigma_6^{abcd}& = f^{abe}d^{cde},\\
 \overline\sigma_7^{abcd}& = f^{ade}d^{bce},\\
 \overline\sigma_8^{abcd}& = f^{cde}d^{abe}.
\end{align}
\end{subequations}
Again, we choose combinations that are symmetric or antisymmetric under exchange of the gluon legs.
In addition, they should also be symmetric or antisymmetric under exchange of the ghost legs.
This yields the basis
\begin{subequations}\label{eq:color_basis_4p}
\begin{align}
 \sigma_1^{abcd}& = -2f^{acd}f^{bde}+f^{abd}f^{cde}, \\
 \sigma_2^{abcd}& = \de^{ab}\de^{cd},\\
 \sigma_3^{abcd}& = \de^{ad}\de^{bc}+\de^{ac}\de^{bd}, \\
 \sigma_4^{abcd}& = -\de^{ad}\de^{bc}+\de^{ac}\de^{bd}, \\
 \sigma_5^{abcd}& = f^{abe}f^{cde},\\
 \sigma_6^{abcd}& = f^{abe}d^{cde},\\
 \sigma_7^{abcd}& = \frac{1}{2}f^{abe}d^{cde}+f^{ade}d^{bce}+\frac{1}{2}f^{cde}d^{abe},\\
 \sigma_8^{abcd}& = f^{cde}d^{abe}.
\end{align}
\end{subequations}
The tensors $\sigma_i$, $i=1,2,3,8$ are symmetric and the others antisymmetric und exchange of $a$ and $b$.
The full symmetry properties are summarized in \tref{tab:color_tensors_symmetries}.
A basis symmetrized with respect to all four legs was derived in ref.~\cite{Eichmann:2015nra}.

If the symmetric structure constant is neglected, the number of independent tensors reduces to five.
One might expect that neglecting $d^{abc}$ leads to problems with an incomplete basis given the relations in \eref{eq:SUN_id_ff}.
However, it was already noted in Ref.~\cite{Driesen:1997wz} that the set $\{\sigma_1,\ldots,\sigma_5\}$ closes under DSE iterations if no symmetric color part from three-point functions is taken into account.
Indeed, this set and the set of the other three tensors, $\{\sigma_6,\ldots,\sigma_8\}$, are orthogonal to each other in $SU(3)$.
Hence, the corresponding parts of vertices decouple from each other.
Furthermore, as explained below, the second set only couples to the symmetric color part in the DSEs of three-point functions.
Thus, it can be completely neglected, since the three-point functions should be color antisymmetric due to charge invariance of QCD \cite{Smolyakov:1980wq,Blum:2015lsa}.
It is thus sufficient to consider only the set $\{\sigma_1, \ldots, \sigma_5\}$.
The possibility that the second set leads to non-zero dressing functions in a self-consistent way is tested below.
$\{\sigma_1, \ldots, \sigma_5\}$ is called the reduced basis in the following.

The full basis for the two-ghost-two-gluon vertex is constructed as the direct product of color and Lorentz space:
\begin{align}
 \rho_{\mu\nu}^{k,abcd}=\sigma_i^{abcd} \tau^j_{\mu\nu}
\end{align}
with $k=k(i,j)=5(i-1)+j$. The vertex is then written as
\begin{align}\label{eq:AAcbc}
 \Gamma^{AA\bar cc,abcd}_{\mu\nu}(p,q;r,s) = g^4 \sum_{k=1}^{40} \rho_{\mu\nu}^{k,abcd} D^{AA\bar cc}_{k(i,j)}(p,q;r,s).
\end{align}
Note that a factor of $g^4$ is put in front.
This accounts for the fact that the lowest diagrams are of this order, because there is no tree-level contribution as for the four-gluon vertex.
Furthermore, the dressing functions are dimensionful, because the vertex is dimensionless, while the Lorentz tensors are dimensionful.

The two-ghost-two-gluon vertex has two distinct DSEs distinguished by which legs are attached to the bare vertices.
The one where this is a gluon ($A$-DSE) contains two-loop diagrams, whereas the one where this is the antighost ($c$-DSE) has a one-loop structure.
Thus, the latter is easier to calculate.
In addition, truncating it is straightforward:
If only the primitively divergent correlation functions are kept, only one diagram is discarded that contains a two-ghost-three-gluon vertex.
The resulting equation, derived with \emph{DoFun} \cite{Alkofer:2008nt,Huber:2011qr}, is depicted in \fref{fig:2gh2gl_DSE}.

The four-ghost vertex is a scalar in Lorentz space.
Thus it features only 8 tensors in total.
The full vertex is written as
\begin{align}\label{eq:cbcbcc}
 \Gamma^{\bar c\bar ccc,abcd}(p,q,r,s) = g^4 \sum_{k=1}^{8} \sigma^{k,abcd} E^{\bar c\bar ccc}_{k}(p,q,r,s).
\end{align}
However, also here the reduced basis is completely decoupled from the other three tensors and thus sufficient.
The truncated DSE of the vertex is shown in \fref{fig:4gh_DSE}.
In the employed truncation only one diagram is dropped that involves a four-ghost-one-gluon vertex.

\section{Results}
\label{sec:results}

The system of ghost-gluon, three-gluon, four-gluon and two-ghost-two-gluon vertices is solved using the propagator results described in Sec.~\ref{sec:props}.
The DSEs for the first three correlation functions are depicted in \fref{fig:ghg-3g-4g_DSEs}.
The DSE for the two-ghost-two-gluon vertex was discussed in Sec.~\ref{sec:4p_funcs}.
The four DSEs are solved in a simple iterative process using a program based on \textit{CrasyDSE} \cite{Huber:2011xc}.
The four-ghost vertex is solved separately using the results for the other vertices, as it does not couple back.

For all vertex functions, a one-momentum approximation is employed, viz., all dressing functions depend only on the average momentum square.
While this approximation has a quantitative effect, it can be expected that the qualitative results hold.
It can even be hoped that the quantitative effect is not that large.
For example, for the three-gluon vertex it is known that it has a very weak angle dependence \cite{Blum:2014gna,Eichmann:2014xya}.

Another approximation is the employed tensor basis for the three- and four-gluon vertices.
They consist only of the respected tree-level tensors.
For the three-gluon vertex, it is known that it is the leading tensor structure \cite{Eichmann:2014xya}.
For the four-gluon vertex, on the other hand, only some other dressing functions have been tested but were also found to be smaller than the tree-level \cite{Cyrol:2014kca}.

\begin{figure*}[tb]
  \includegraphics[width=0.48\textwidth]{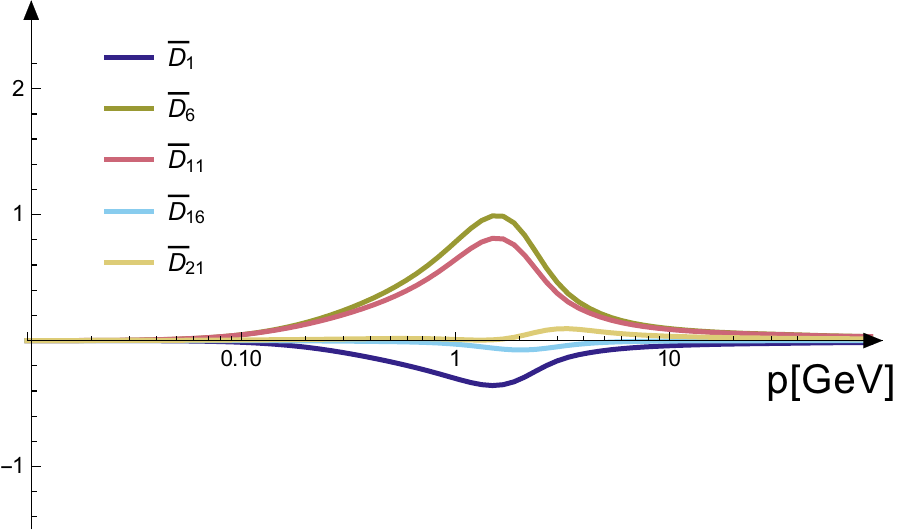}\hfill
  \includegraphics[width=0.48\textwidth]{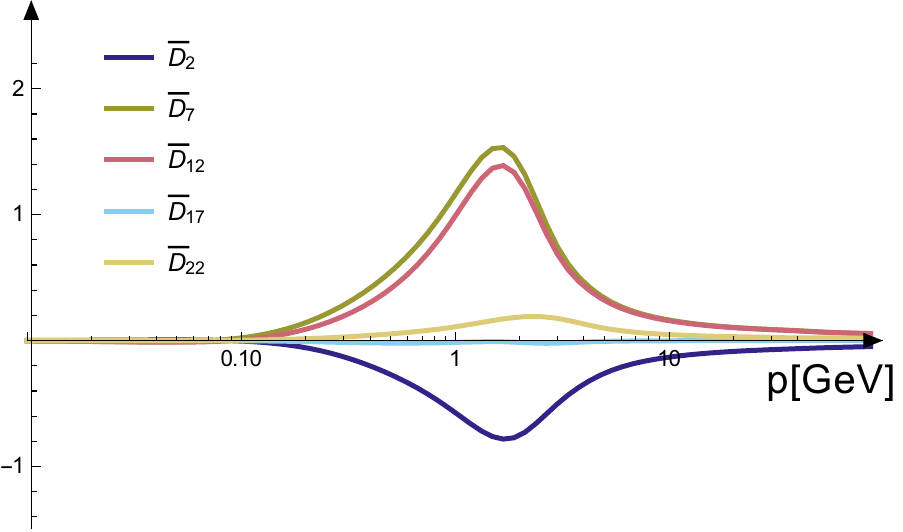}\\
  \vskip5mm
  \includegraphics[width=0.48\textwidth]{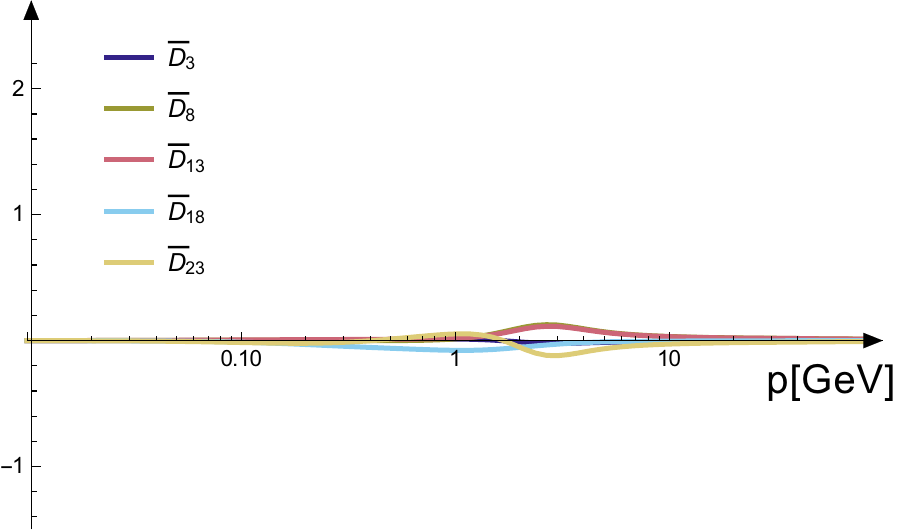}\hfill
  \includegraphics[width=0.48\textwidth]{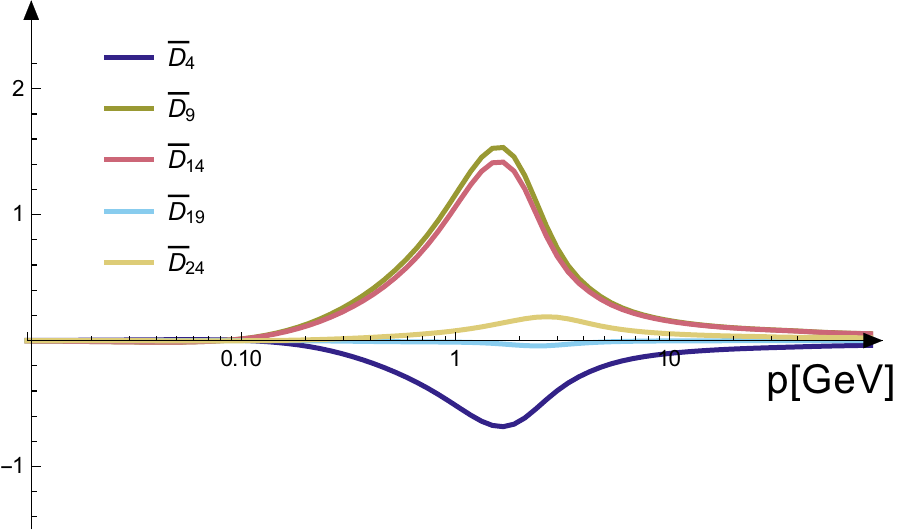}
  \vskip5mm
  \includegraphics[width=0.48\textwidth]{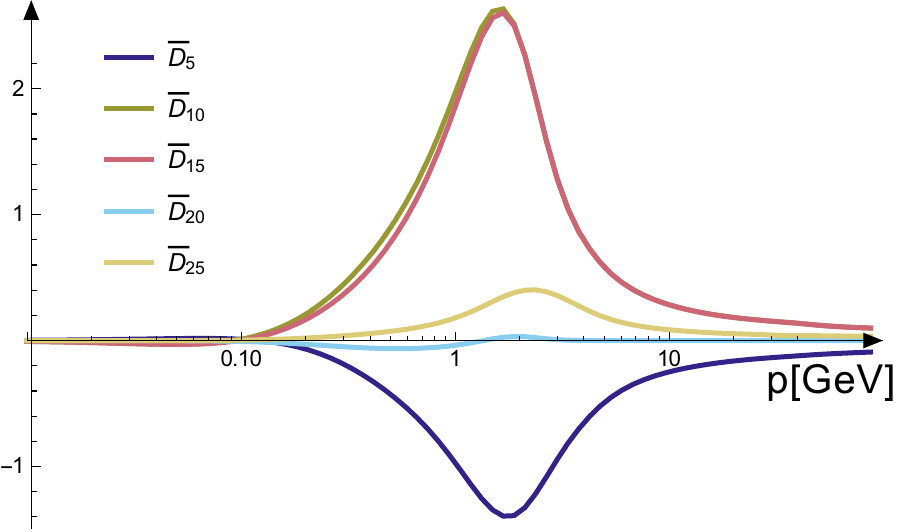}\hfill
  \begin{center}
  \caption{\label{fig:res_2gh2gl}Results for the two-ghost-two-gluon vertex.
  Each plot shows the results corresponding to one Lorentz tensor.
  The dressing functions related to the color tensors $\sigma_6$, $\sigma_7$ and $\sigma_8$ are compatible with zero.}
 \end{center}
\end{figure*}

\subsection{Two-ghost-two-gluon vertex}
\label{sec:res_2gh2gl}

As mentioned in Sec.~\ref{sec:4p_funcs}, the color part of the two-ghost-two-gluon vertex is split into two orthogonal parts of which only one couples to the other correlation functions.
To test the possibility whether the decoupled part creates a nonzero solution self-consistently, it was calculated as well using as ansatz a power law based on the dimensionality of the dressing functions.
The iteration process reduced the magnitude of the dressing function in each step and no stable solution was found.
Thus, provided that the neglected diagram does not change that or the iteration methods misses a possible solution, this part of the vertex is indeed zero.

Results for the dressing functions corresponding to the reduced color basis are shown in \fref{fig:res_2gh2gl}.
For the images, the dressing functions $D^{AA\bar cc}$ were rendered dimensionless: $D^{AA\bar cc}_k\rightarrow \overline D_k$.
This also reflects better their actual impact in DSEs, where they enter with dimensionful basis tensors.
There are two remarkable findings.
First, the dressings corresponding to the Lorentz tensor $\tau^3$ are small.
Second, all dressing corresponding to the color tensors $\sigma_4$ and $\sigma_5$ are small.
These two are the ones anti-symmetric under exchange of the two gluon legs.
The UV behavior is found as expected to vanish with a power law behavior.

\subsection{The four-ghost vertex}

\begin{figure}[tb]
  \includegraphics[width=0.48\textwidth]{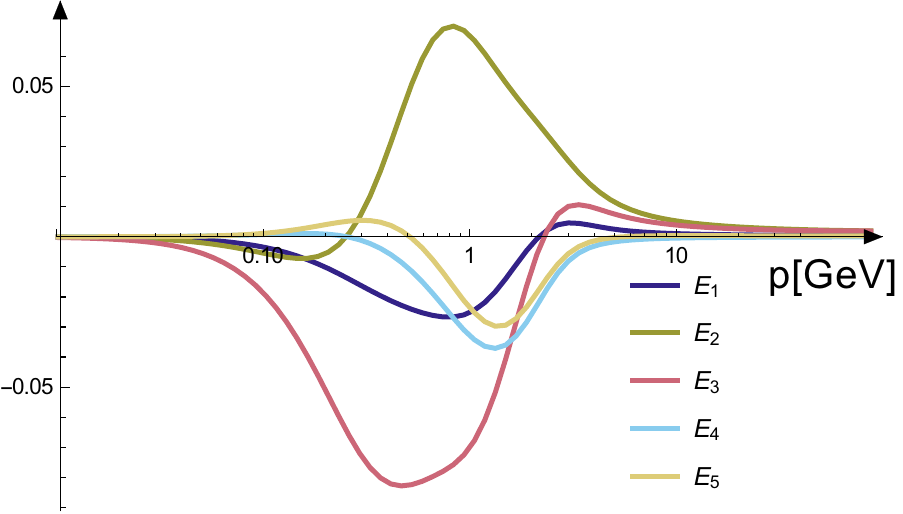}
  \begin{center}
  \caption{\label{fig:res_4gh}Results for the four-ghost vertex.
  The dressing functions $E_6$, $E_7$ and $E_8$ are compatible with zero.}
 \end{center}
\end{figure}

Results for the four-ghost vertex are shown in \fref{fig:res_4gh}.
As for the two-ghost-two-gluon vertex, no non-zero solution for the dressing functions corresponding to the color tensors $\sigma_5$, $\sigma_7$ and $\sigma_8$ was found.
The other five dressing functions are rather small, at the order of the small dressing functions of the two-ghost-gluon vertex.

\begin{figure}[tb]
 \includegraphics[width=0.45\textwidth]{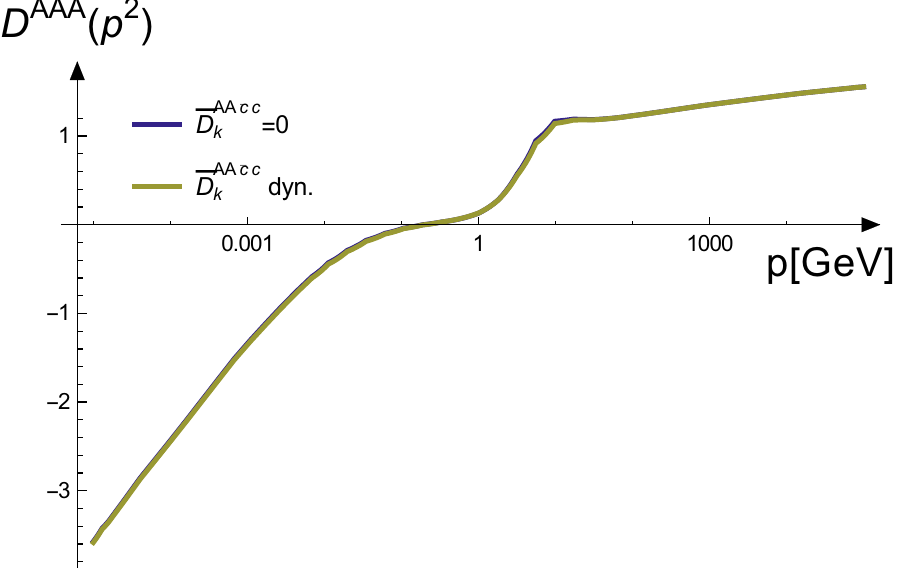}
 \caption{Comparison of the three-gluon vertex with and without the two-ghost-two-gluon vertex.}
 \label{fig:comp_3g}
\end{figure}

\subsection{Influence on three-point functions}

The two-ghost-two-gluon vertex appears in the three-gluon vertex and the ghost-gluon vertex DSEs in diagrams called swordfish diagrams.
As the two-ghost-two-gluon vertex, the ghost-gluon vertex has two different DSEs, the $A$- and $c$-DSEs.
The $A$-DSE is the more complicated one of the two containing also two-loop diagrams, see \cite{Alkofer:2008nt} for the full equation.
The $c$-DSE, on the other hand, contains only 4 diagrams, see \fref{fig:ghg-3g-4g_DSEs}: the tree-level diagram, two triangle diagrams and the swordfish diagram with the two-ghost-two-gluon vertex. 

In the three-gluon vertex DSE, the two-ghost-two-gluon vertex is contracted with a bare three-gluon vertex which is proportional to the antisymmetric structure constant $f^{abc}$.
In addition, also the full vertex contains only $f^{abc}$ as long as parity is conserved \cite{Smolyakov:1980wq}.
Thus, only tensors antisymmetric with respect to $a\leftrightarrow b$ and $c\leftrightarrow d$ can contribute.
There are only two with these properties, $\sigma_4$ and $\sigma_5$.
Exactly these are the tensors that were found to be the smallest ones in Sec.~\ref{sec:res_2gh2gl} and one can thus expect only a minor quantitative influence on the three-gluon vertex.
Indeed, an explicit comparison of calculations with and without the two-ghost-two-gluon vertex shows that the difference the ghost swordfish diagram makes is not visible with the eye, see \fref{fig:comp_3g}.

\begin{figure}[tb]
 \includegraphics[width=0.45\textwidth]{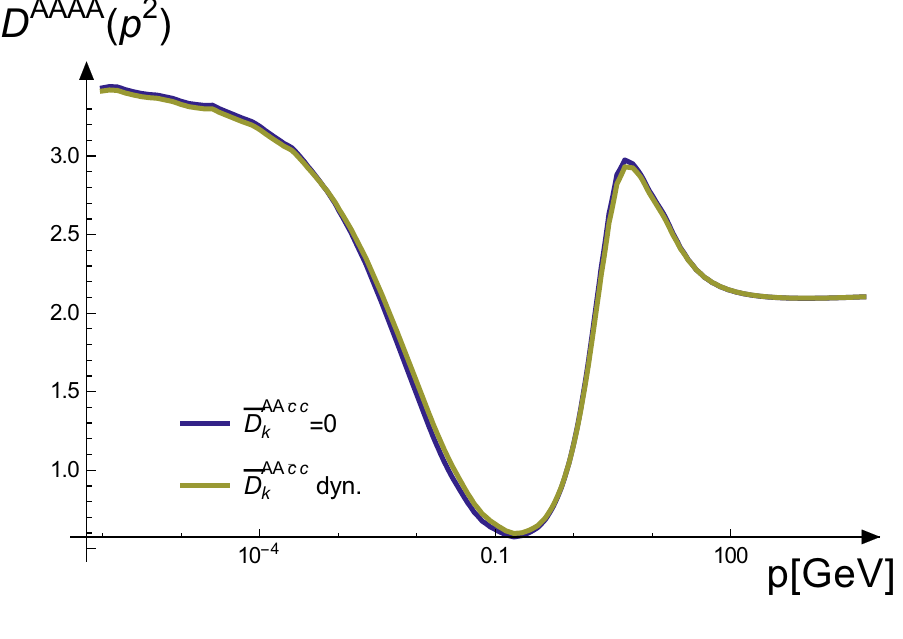}
 \caption{Comparison of the four-gluon vertex with and without the two-ghost-two-gluon vertex.}
 \label{fig:comp_fg}
\end{figure}

Doing the same comparison for the ghost-gluon vertex shows that in this case there is a visible difference.
However, compared to the overall amplitude it is very small, namely $1.7\%$ at most.
The influence of the two-ghost-two-gluon vertex on the four-gluon vertex is found to be even smaller, as can be seen in \fref{fig:comp_fg}.

\begin{figure}[tb]
 \includegraphics[width=0.45\textwidth]{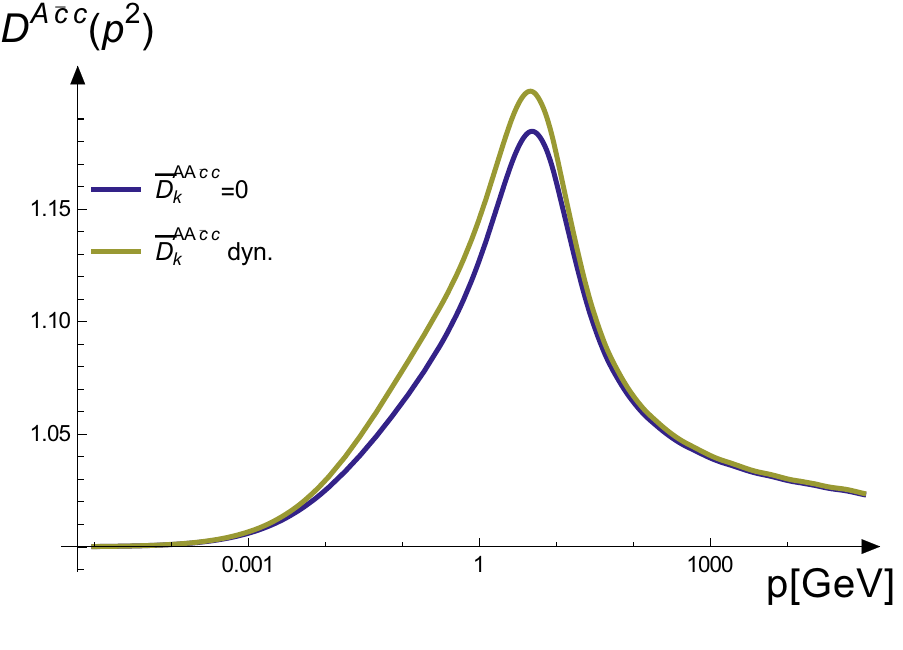}
 \caption{Comparison of the ghost-gluon vertex with and without the two-ghost-two-gluon vertex.}
 \label{fig:comp_ghg}
\end{figure}

The four-ghost vertex does not appear in either of the other equations considered here.
It should be stressed that this is a genuine feature and not due to the employed truncation.
The lowest equation where it appears is the $A$-DSE of the ghost-gluon vertex.
It also appears in the $A$-DSE of the two-ghost-two-gluon vertex.
Given the smallness of the dressing functions of the four-ghost vertex, no big influence of this vertex should be expected.
However, those equations are not used here as it is advantageous also for other reasons to use the corresponding $c$-DSEs.

\section{Conclusions}
\label{sec:conclusions}

The two-ghost-two-gluon vertex and the four-ghost vertex were studied for the first time with functional equations.
Their full (transverse) bases were employed.
In color space, three tensors decouple completely from the equations themselves and also from the three-point functions.
The only self-consistent solution that was found for this part was compatible with zero.
The other five dressing functions of the four-ghost vertex are very small and would not have any sizable impact in other equations.
However, choosing for correlation functions with a ghost leg always the $c$-DSE, viz., the version of the DSE where only the ghost-gluon vertex appears bare, the four-ghost vertex does not contribute to any other $n$-point function for $n\leq4$.

The two-ghost-two-gluon vertex, on the other hand, appears in all three- and four-point functions.
The solution for this vertex reveals that all dressing functions proportional to two of the five relevant color tensors are very small.
In total only twelve dressing functions have some strength.
Interestingly, these dressings all do not contribute in the three-gluon vertex DSE due to their color structure.
Thus, there is basically no effect on this vertex.
In the DSE of the the ghost-gluon vertex, which was solved without truncation, a small effect at the order of $2\%$ was found.

It is interesting that in the propagator equations of the FRG the two-ghost-two-gluon vertex appears in tadpole diagrams.
Due to the color structure of these diagrams, only the dressing functions which are not small survive.
Thus, the inclusion of this vertex could have an impact, but, as the example of the ghost-gluon vertex shows, most likely only a small one.
For DSEs, such an influence can only enter indirectly via primitively divergent correlation functions.

The results of this work provide more evidence that the tower of functional equations can be truncated in a meaningful way that allow a self-contained solution.
First of all, many dressing functions of the two non-primitively divergent correlation functions calculated here were found to be small.
Second, their impact on lower correlation functions was found to be very small as well.

This study focused on a proper description of two vertices not investigated previously.
To complement this study, also for the three- and four-gluon vertices calculations with their full tensor bases should be done.
While this will not change the direct impact of the two-ghost-two-gluon vertex on the three-gluon vertex as the vanishing of the relevant contributions happens in color space,
indirect effects, e.g., via the four-gluon vertex cannot be excluded.

\section*{Acknowledgments}

I thank Mario Mitter for useful discussions. HPC Clusters at the University of Graz were used for the numerical computations.
Support by the FWF (Austrian science fund) under Contract No. P27380-N27 is gratefully acknowledged.

\appendix

\section{$SU(3)$ color identities}
\label{sec:color}

The $SU(N)$ algebra is defined by
\begin{align}
 [T^a,T^b]=i \,f^{abc}T^a,
\end{align}
where the $T^a$ are the generators.
The normalization
\begin{align}
 \text{Tr} \left( T^a T^b\right)=\frac{1}{2} \de^{ab}
\end{align}
is used.
$f^{abc}$ is the antisymmetric structure constant.
A symmetric structure constant $d^{abc}$ can be defined via the anticommutator:
\begin{align}
 \{T^a,T^b\}=\frac{1}{N}\de^{ab}+d^{abc}T^c.
\end{align}

Some relations of products of structure constants are well known, e.g., $f^{amn}f^{bmn}=N \de^{ab}$ or $f^{amn}f^{bno}f^{com}=N/2 f^{abc}$.
When symmetric structure constants are involved, corresponding identities are more difficult to find.
Here is a collection of those used in this work for general $N$:
\begin{subequations}
 \begin{align}
  f^{amn}d^{bmn}&=0,\\
  d^{amn}f^{bno}f^{com}&= -\frac{N}{2} d^{abc},\\
  d^{amn}d^{bno}f^{com}&=\left(\frac{2}{N}-\frac{N}{2}\right) f^{abc},\\
  d^{amn}d^{bno}d^{com}&=\left(\frac{N}{2}-\frac{6}{N}\right) d^{abc}.
\end{align}
For $N=3$, the following identities were used:
\begin{align}
  &f^{amn}f^{bno}f^{cop}f^{dpm}=-\frac{N}{3} f^{abe}f^{cde}-\frac{N}{6} f^{ace}f^{dbe}\nonumber\\
  &\quad+\frac{5N^2}{6(N^2+1)}\left(\de^{ab}\de^{cd}+\de^{ac}\de^{bd}+\de^{ad}\de^{bc}\right),\\
  &d^{amn}f^{bno}f^{cop}f^{dpm}=\frac{N}{4} \left(f^{bce} d^{ade} + f^{bde} d^{ace} + f^{cde} d^{abe}\right),\\
  &f^{amn}f^{bno}d^{cop}d^{dpm}=\frac{5N^2}{18(N^2+1)}\left(\de^{ad}\de^{bc}+\de^{ac}\de^{bd}\right)\nonumber\\
  &\quad- \frac{19N^2-36}{18(N^2+1)}\de^{ab}\de^{cd}+\frac{2(N^4-2N^2-18)}{9N(N^2+1)} f^{abe}f^{cde} \nonumber\\
  &\quad- \frac{108-11N^2+N^4}{18N(N^2+1)}f^{ace}f^{dbe},\\
  &d^{amn}d^{bno}d^{cop}f^{dpm}=-\frac{1}{4} N d^{cde}f^{abe} - \frac{2}{N} d^{bce}f^{ade}\nonumber\\
  &\quad- \frac{-8 + N^2}{4 N}d^{abe}f^{cde},\\
  &d^{amn}d^{bno}d^{cop}d^{dpm}=-\frac{-1080 + 396 N^2 - 49 N^4}{54 N^2 (1 + N^2)} \de^{ad}\de^{bc} \nonumber\\
  &\quad- \frac{-864 + 612 N^2 - 49 N^4}{ 54 N^2 (1 + N^2)} \de^{ac}\de^{bd} \nonumber\\
  &\quad- \frac{-1080 + 288 N^2 - 37 N^4}{ 54 N^2 (1 + N^2)} \de^{ab}\de^{cd}\nonumber\\
  &\quad- \frac{-36 - 85 N^2 + 11 N^4}{ 27 N (1 + N^2)} f^{abe}f^{cde}\nonumber\\
  &\quad- \frac{72 - 151 N^2 + 17 N^4}{54 N (1 + N^2)} f^{ace}f^{dbe}.
 \end{align}
\end{subequations}
The expression for four $f$'s in the case of $N=2,3$ agrees with that of Ref.~\cite{Cyrol:2014kca}.
All identities were tested numerically for $SU(3)$.
As explained in Sec.~\ref{sec:4p_funcs}, an additional independent color tensor in the basis of tensors with four indices appears for $N>3$ and would thus need to be taken into account for deriving the identities above.

\bibliographystyle{utphys_mod}
\bibliography{literature_nonPrimDivVertices}

\end{document}